\documentclass[nohyperref]{article}

\usepackage{microtype}
\usepackage{graphicx}
\usepackage{subfig}
\usepackage{booktabs} 

\usepackage{amsmath,amsfonts,bm}

\def\eqref#1{equation~\ref{#1}}
\def\Eqref#1{Equation~\ref{#1}}

\def\1{\bm{1}}

\def\vtheta{{\bm{\theta}}}

\def\vw{{\bm{w}}}

\def\mX{{\bm{X}}}
\def\mY{{\bm{Y}}}

\DeclareMathAlphabet{\mathsfit}{\encodingdefault}{\sfdefault}{m}{sl}
\SetMathAlphabet{\mathsfit}{bold}{\encodingdefault}{\sfdefault}{bx}{n}

\newcommand{\E}{\mathbb{E}}

\newcommand{\KL}{D_{\mathrm{KL}}}

\DeclareMathOperator*{\argmin}{arg\,min}

\usepackage{hyperref}
\usepackage{url}
\usepackage{subfig}
\usepackage{nicematrix} 
\usepackage{booktabs}
\usepackage{placeins}
\usepackage{wrapfig}
\usepackage[ruled,vlined]{algorithm2e}

\newcommand{\norm}[1]{\left\lVert#1\right\rVert}

\usepackage[accepted]{icml2022}

\usepackage{amsmath}
\usepackage{amssymb}
\usepackage{mathtools}
\usepackage{amsthm}

\usepackage[capitalize,noabbrev]{cleveref}

\theoremstyle{plain}

\theoremstyle{definition}

\theoremstyle{remark}

\usepackage[textsize=tiny]{todonotes}

\newcommand{\nt}[1]{#1} 
\newcommand{\mm}[1]{#1}

\icmltitlerunning{BNNs and Fluid Simulations}

\begin{document}

\twocolumn[
\icmltitle{Leveraging Stochastic Predictions of Bayesian Neural Networks \\ for Fluid Simulations}



\icmlsetsymbol{equal}{*}

\begin{icmlauthorlist}
\icmlauthor{Maximilian Mueller}{tueb,ipp}
\icmlauthor{Robin Greif}{ipp,tum}
\icmlauthor{Frank Jenko}{ipp,tum}
\icmlauthor{Nils Thuerey}{tum}
\end{icmlauthorlist}

\icmlaffiliation{tueb}{Department of Computer Science, University of Tübingen, Tübingen, Germany}
\icmlaffiliation{ipp}{Max Planck Insitute for Plasma Physics, Munich, Germany}
\icmlaffiliation{tum}{Department of Computer Science, TU Munich, Germany}

\icmlcorrespondingauthor{Maximilian Mueller}{maximilian.mueller@wsii.uni-tuebingen.de}


\vskip 0.3in
]



\printAffiliationsAndNotice{} 

\begin{abstract}
We investigate uncertainty estimation and multimodality via the non-deterministic predictions of Bayesian neural networks (BNNs) in fluid simulations. To this end, we deploy BNNs in three challenging experimental test-cases of increasing complexity: 
We show that BNNs, when used as surrogate models for steady-state fluid flow predictions, provide accurate physical predictions together with sensible estimates of uncertainty.
Further, we experiment with perturbed temporal sequences from Navier-Stokes simulations and evaluate the capabilities of BNNs to capture multimodal evolutions. While our findings indicate that this is problematic for large perturbations, our results show that the networks learn to correctly predict high uncertainties in such situations.
Finally, we study BNNs in the context of solver interactions with turbulent plasma flows. We find that BNN-based corrector networks can stabilize coarse-grained simulations and successfully create multimodal trajectories.
\end{abstract}

\vspace{-0.1cm}
\section{Introduction}
Even though Bayesian neural networks (BNNs) have been studied for a long time in the Machine Learning community \citep{hinton_keeping_1993,mackay_practical_1992}, they have only recently received increased attention in the wild. While conventional, non-Bayesian deep learning techniques, that have mostly been used in fluid simulation setups, provide point estimates, BNNs allow to obtain stochastic predictions, since they learn a distribution over the network's weight parameters. Exploring to which extent those stochastic predictions can be exploited in the context of fluid simulations is the central goal of this work. \nt{In particular, we identify and investigate two use-cases of a combination of BNNs with fluid simulations: uncertainty estimation and multi-modal synthesis.}

A central motivation for the use of BNNs is the estimation of uncertainty \citep{mackay_bayesian_1992}. 
In the context of fluid simulations, it is an open question whether neural networks can successfully provide sensible uncertainty estimates.
This includes uncertainty location on the one hand: If a neural network is for instance deployed as surrogate model to a physical solver, \mm{the uncertainty locations in the predictions should correspond to regions that are harder to predict, \nt{e.g., more turbulent locations}.} On the other hand, a key challenge is to relate the predictive performance to the uncertainty that comes along with BNN predictions quantitatively. In practice, this can be done by comparing a suitable measure of predictive performance, such as the mean absolute error (MAE), to a measure of uncertainty, e.g., the standard deviation over repeated predictions. We then expect the standard deviation to correlate with the mean absolute error. 
Ideally, this way the model can communicate when it is uncertain about a prediction, or when the learning process failed to converge. 

Further, fluid models typically provide a deterministic description of an inherently chaotic and stochastic process \citep{pope_2000}.
Many initial conditions lead to bifurcation points where epsilon changes of the flow state can lead to fundamentally different solutions over time \citep{Ko2008SensitivityConditions}, e.g., a vortex turning left or right.
Since conventional neural networks typically act as deterministic predictors that provide point estimates, they are limited in describing such setups. It is therefore interesting to investigate if the stochasticity of the BNN predictions can be exploited in order to resemble multimodal solutions. One particular case of interest is plasma physics transport modelling \citep{balescu_aspects_2005},  one of the key challenges on the path to a practical fusion device \citep{freidberg_plasma_2008}. There, transport is driven by micro-instabilities and the resulting systems are highly turbulent. Consequently, minor changes in the initial conditions can lead to very different solution states after a short period of time, creating a particularly relevant setup to study multimodality. 

\newpage
In the following, we assess the performance of BNNs on these tasks in three test-cases. We show that a trained BNN  produces meaningful uncertainty estimates for complex fluid simulation scenarios, like Reynolds-averaged Navier-Stokes flow around airfoils, and perturbed buoyancy-driven Navier-Stokes flow. In addition, our experiments demonstrate that BNNs successfully generate varied predictions  when working in conjunction with a numerical simulator in the plasma turbulence setup. 

\section{Related Work}

Leveraging data-driven methods in the context of PDE models has been a long-standing goal \citep{brunton_discovering_2016,bindal_equation-free_2006,crutchfield_equations_1987}. In the past years, the focus has shifted towards deep-learning approaches. Those were successfully applied to a wide variety of tasks, such as 
deep learning for reduced models with Koopman theory \citep{li_learning_2020,morton_deep_2018},
identifying model equations \citep{raissi_hidden_2018,long_pde-net_2018},
and learned discretizations \citep{bar-sinai_learning_2019}.
Additionally, \citet{tompson_accelerating_2017} investigated unsupervised learning of corrections while \citet{sirignano_dgm_2018} used physics-informed methods by deploying PDE-based loss functions. 
Other research focused on efficient simulations by learning conservation laws \citep{cranmer_lagrangian_2020,greydanus_hamiltonian_2019}, or aimed at correcting iterative solvers \citep{hsieh_learning_2019}. 
Turbulence modelling received particular attention, e.g., from \citet{beck_deep_2019},  \citet{tracey_machine_2015} and \citet{Novati2021AutomatingLearning}. 

Convolutional neural networks, which we will likewise use in our experiments, were used in the context of flow problems as the basis 
for generative models \citep{chu_data-driven_2017,kim_deep_2019}, or for
corrector models \citep{um_iquid_2018,thuerey_deep_2020}.
Further, the recent development of geometric deep learning approaches \citep{bronstein_geometric_2021} also impacted fluid flow problems: Mesh-based methods \citep{pfaff_learning_2020}, graph neural networks \citep{sanchez-gonzalez_learning_2020} and continuous convolutions \citep{ummenhofer_lagrangian_2020} were deployed successfully. 

Differentiable components and differentiable programming have been leveraged by a variety of recent works \citep{amos_optnet_2017,innes_differentiable_2019,hu_difftaichi_2020,chen_neural_2019}. 
These approaches enable an end-to-end training which was shown to have advantages in rigid body control \citep{Belbute-Peres2018} and advection-diffusion systems \citep{yin2021augmenting}.
Within the scope of our work, we will use the differentiable simulator \textit{PhiFlow} \citep{holl_learning_2020}, in particular in a tight integration with neural networks, as suggested by \citet{um_solver---loop_2021}. 

Early contributions to Bayesian networks can be attributed to \citet{mackay_bayesian_1992,mackay_practical_1992}, \citet{hinton_keeping_1993} and \citet{neal_bayesian_1996}. In this paper, we will use Bayesian networks based on variational inference \citep{kingma_variational_2015}, taking advantage of reparametrization techniques \citep{kingma_auto-encoding_2014,blundell_weight_2015} and Monte-Carlo-based methods \citep{GalDropoutPaper,graves_practical_nodate,welling_bayesian_2011}.
In the wild, BNNs have mostly been deployed for segmentation tasks \citep{badrinarayanan_segnet_2016}:  \citet{labonte_we_2020} obtained 3D geometric uncertainties for CT scans, \citet{kwon_uncertainty_nodate} performed image segmentation on biomedical data, \citet{deodato_bayesian_2019} on cellular images and \citet{mcclure_knowing_2019} on brain segmentation tasks. More recently, \citep{BayesianPlanets2021} leveraged BNNs to predict the dissolution of compact planetary systems. 
Even though uncertainty quantification has been a long-standing topic for computational fluid dynamics \citep{roache_quantification_1997}, the use of BNNs in this area remains largely unexplored - with a few exceptions like \citet{sun2020physicsconstrained}, who deployed a physics-constrained Bayesian network to reconstruct fluid flow from sparse and noisy data. In contrast to their work, we do not constrain our networks and investigate a broader class of problems and use-cases of BNNs.

\begin{figure*}[!tb]
    \centering
    \includegraphics[width=1.\textwidth]{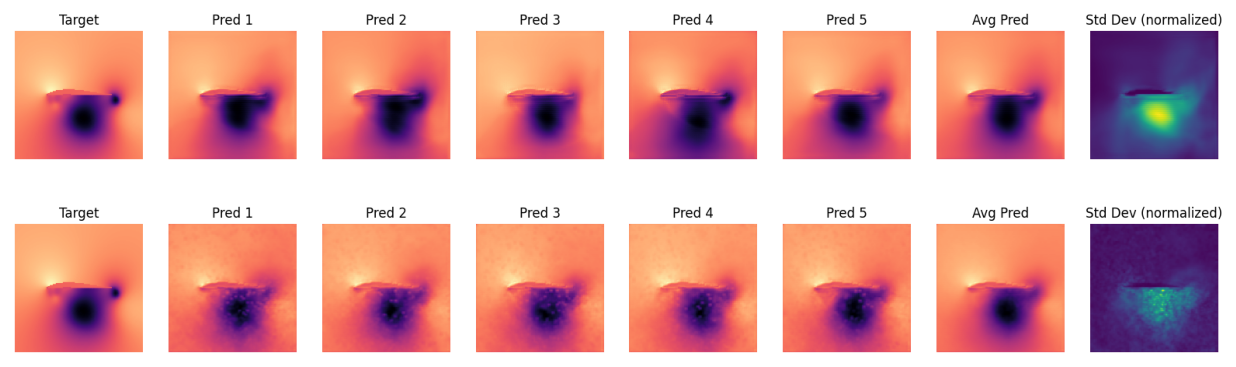}
    \caption{Repeated samples from BNN with spatial dropout (top) and conventional dropout (bottom) for a specific (hard) test case. The individual predictions of the BNN with spatial dropout are smoother and show larger variations compared to the conventional dropout case, where the difference between predictions is on a smaller scale.}
    \label{fig:spatial_vs_normal_repeated}
\end{figure*}

\section{Background: Bayesian Neural Networks}
Bayesian Neural Networks provide stochastic predictions by incorporating the Bayesian paradigm into deep learning. The network weights $\vw$ are thought to follow a prior distribution $p(\vw)$, which is updated to the posterior distribution $p(\vw|\mX,\mY)$ after observing the data consisting of inputs $\mX$ and targets $\mY$. In this work, we used two variants of BNNs. Both stem from variational inference \citep{kingma_variational_2015} and aim at minimizing the KL-divergence between true and approximate posterior, or equivalently maximize the evidence lower bound (ELBO): 
\begin{equation}\label{eq:ELBO}
    \E_{\vw\sim q_{\vtheta}} [ \log p(\mY|\vw,\mX) ] - \frac{1}{\lambda}\KL \left(q_{\theta}(\vw) \Vert p(\vw)\right)
\end{equation}
where $q_{\vtheta}(\vw)$ is the approximate posterior distribution over the weights, which is parametrized by $\theta$. \mm{$\KL$ denotes the KL-divergence and }$\lambda\geq 1$ is a scaling factor that was empirically shown to improve the performance of BNNs (see e.g. \citet{wenzel_how_2020}).  We choose our noise model such that the log-probability can be written as the mean absolute error, which has shown to lead to good performance in 2d fluid settings \citep{deepflowReynoldsThuerey}, or the mean squared error. Intuitively, the two terms in \eqref{eq:ELBO} have opposite goals: Maximizing the expected log-probability encourages the variational distribution to fit the data well. Minimizing the KL-divergence, in contrast, forces the approximate posterior distribution to stay close to the prior, which penalizes complex distributions and can be seen as a form of regularization. The scaling factor $\lambda$ intuitively assigns different weight to those goals: For $\lambda\to1$ we have a Bayesian network where the log-likelihood and the KL-term receive equal weight. For $\lambda \to \infty$ the second term disappears, and the optimization objective turns into the negative log-likelihood. In order to estimate \eqref{eq:ELBO}, and in particular derivatives thereof, sub-sampling and Monte-Carlo techniques are typically leveraged together with a reparametrization trick \citep{kingma_auto-encoding_2014}, that allows to backpropagate gradients through distributions. One particular stochastic estimator we use in this work is the \textit{flipout} estimator \citep{wen_flipout_2018}, which computes decorrelated stochastic gradient estimates of \eqref{eq:ELBO} with few perturbation samples. Additionally, we leverage the work of \citet{GalDropoutPaper}, who showed that under mild conditions, conventional neural networks that were trained with dropout regularization, can be seen as a form of Bayesian neural networks. A Dropout layer randomly sets input units to 0 with the specified rate. Spatial Dropout, which we will also use in this work, sets entire feature-maps to 0. \mm{For both variants, obtaining stochastic predictions according to the posterior distribution is as simple as extending dropout to the prediction phase.} In all considered BNNs, the marginal prediction can be obtained by computing the mean of repeated forward passes of a given input. Likewise, the standard deviation over the repeated samples can be seen as a measure of uncertainty. 

\section{Experiments}\label{sec:experiments}

We investigate three scenarios of increasing complexity: (1) A static setup, where the learning goal is to infer steady-state Reynolds-averaged Navier-Stokes (RANS) solutions around airfoils. (2) A perturbed buoyancy-driven Navier-Stokes (NS) flow for which the time evolution is taken into account, and (3) turbulent Hasegawa-Wakatani simulations with a tightly integrated BNN. 

In the following, we will explain the experimental setup and discuss the corresponding results for each of the three cases. 

\subsection{Reynolds-averaged Navier-Stokes flow}

Previous work has successfully used conventional neural networks to infer RANS solutions around airfoils \citep{deepflowReynoldsThuerey}. We follow this experimental setup, but instead deploy a dropout Bayesian neural network as surrogate model. Thus, we investigate if BNNs are capable of obtaining similar results and can provide sensible uncertainty information.

\begin{table*} 
        \centering
        \caption{RANS-Flow Performance}
        \label{tab:my_label}
        \begin{NiceTabular}{@{}c cccc cccc@{}}
        \toprule 
        \Block{2-1}{} &  \Block{1-4}{Dropout} &&&& \Block{1-4}{Spatial dropout} \\[0.01cm] 
        
        &  0.01  & 0.05  & 0.1 & 0.25  & 0.01  & 0.05   & 0.1  & 0.25\\\cmidrule(lr){2-5}\cmidrule(lr){6-9}
        {\textbf{Non-Bayesian MAE} $\times100$}  & 0.70 & 0.70 & {0.75} & 0.98 & 0.60 & 0.70 & \bf{0.73} & \bf{0.96}  \\[0.1cm]
        {\textbf{BNN MAE-avg} $\times100$}  & \bf{0.69} & \bf{0.68} & \bf{0.64} & \bf{0.72} & \bf{0.59} & 0.70 & 0.77 & 0.97  \\[0.1cm]
        {\textbf{BNN MAE-std} $\times100$} & 0.26 & 0.45 & 0.58 & 0.78 & 0.33 & 0.60 & 0.78 & 1.10
        \end{NiceTabular}\label{tab:RansPerformance}
\end{table*}
\textbf{Setup.}
We use the open-source code \textit{OpenFOAM}, which solves a one equation turbulence model (Spalart-Allmaras), to generate ground truth data for training. We consider a range of Reynolds numbers $Re=[0.5,5]\times10^{6}$, incompressible flow and angles of attack in a range of $[-22.5^{\circ}, +22.5^{\circ}]$. With this, we simulate velocity and pressure distributions of flows around 1505 different airfoil shapes from the UIUC database \citep{UiucAirfoilDataSite}. Following \citet{deepflowReynoldsThuerey}, where a more detailed explanation of the data-generating process is available, we encode freestream conditions and airfoil shape in a $128\times 128\times 3$ grid, denoting 3 fields, each at $128\times128$ resolution: \mm{The first field} is a mask of the airfoil shape, the other two $x$- and $y$- velocity components, respectively. The output data sets have the same size, but now the first channel describes the pressure $p$, whereas the other two channels still contain $x$- and $y$- velocity components of the desired RANS solution. 
We normalize with respect to the freestream velocity \mm{by: (1) dividing the target velocity $\mathbf{v}_{o}$ by the magnitude of the input velocity $\mathbf{v}_{i}$, $\Tilde{\mathbf{v}}_{o}={\mathbf{v}}_{o}/|{\mathbf{v}}_{i}|$, (2) the target pressure ${{p}}_{o}$ by the square of the input velocity, $\Tilde{{p}}_{o}={{p}}_{o}/|{\mathbf{v}}_{i}|^{2}$ and (3) removing the mean pressure from each solution.} This learning problem corresponds to a regular supervised setup and can be formalized as minimizing
\begin{equation}
    \E_{\vw\sim q_{\vtheta}} \left[ \sum_{i=1}^{N}\norm{{f}(\mathbf{x}_{i}|\vw)-\mathbf{y}_{i}}_{1} \right] - \frac{1}{\lambda}\KL 
\end{equation}
with respect to $\mathbf{\theta}$. $f$ is a Bayesian neural network whose weights $\mathbf{w}$ are sampled from the approximate posterior distribution. $\mathbf{x}_{i}$ and $\mathbf{y}_{i}$ are the $i^{\text{th}}$ input and target, respectively, each consisting of three $128\times128$ grids. For readability, we omitted the arguments of $\KL$, which are the prior and approximate posterior distributions like in \eqref{eq:ELBO}. We train a U-Net with Monte-Carlo dropout (both spatial dropout and conventional dropout) for 100 epochs with a learning rate of 0.006, learning rate decay and the \textit{Adam} optimizer \citep{kingma_adam_2017}. Details of the neural network architecture can be found in the appendix.

\textbf{Results.}
We find that the considered BNNs show similar performance to the non-Bayesian networks of \citet{deepflowReynoldsThuerey} in terms of mean absolute error. Table \ref{tab:RansPerformance} illustrates that in our experiments, most BNNs are even slightly superior to their non-Bayesian counterparts. \mm{Since the non-Bayesian networks are likewise trained with dropout, we hypothesize that this is not caused by stronger regularization in BNNs.} Instead, we think that the BNN posterior can successfully capture many compelling but different solutions, which are combined in an ensemble-like manner during marginalization \citep{BayesianDLWilson}. 

\begin{figure*}[htb]
    \centering
    \includegraphics[width=.48\textwidth]{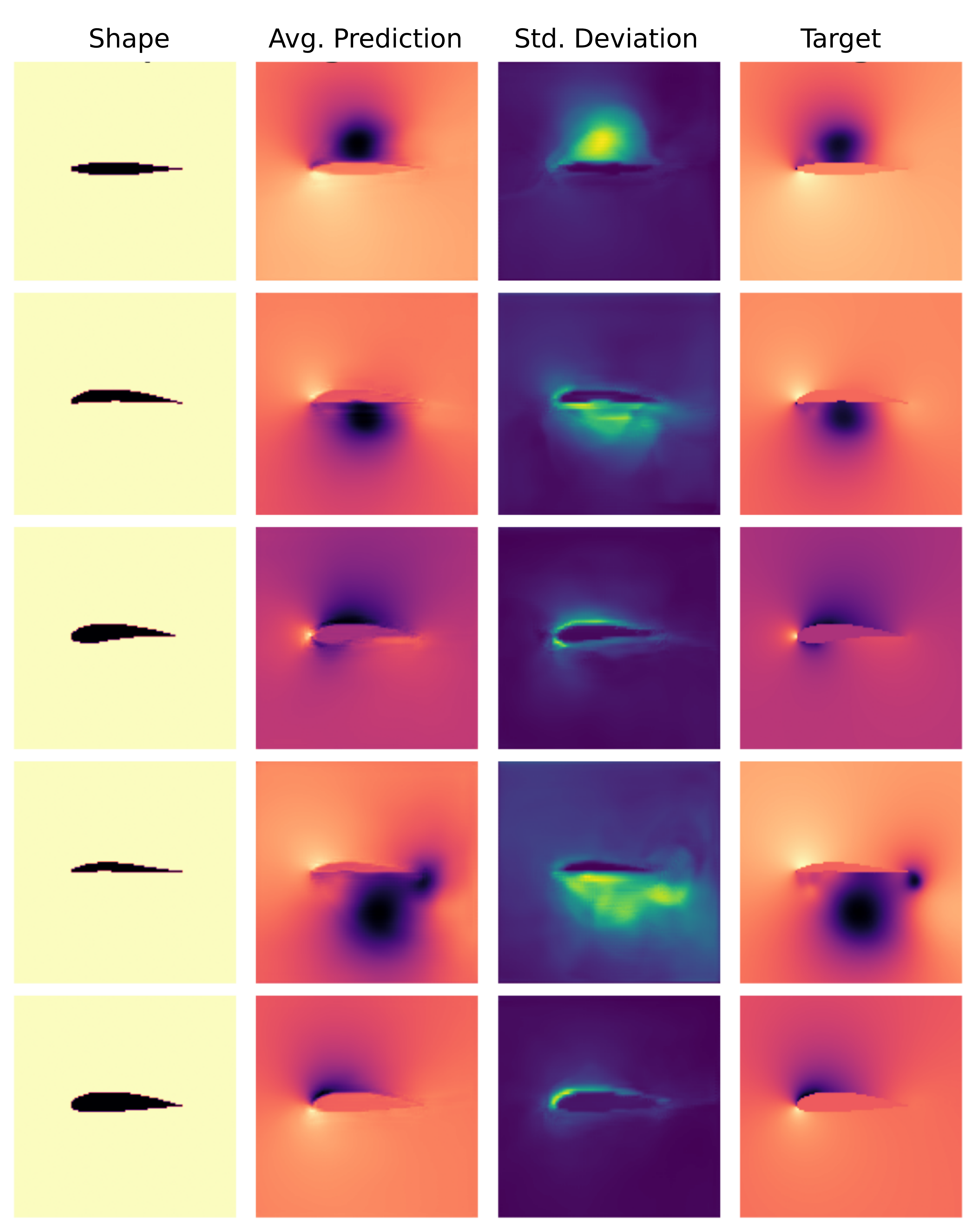}
    \quad 
    \includegraphics[width=.48\textwidth]{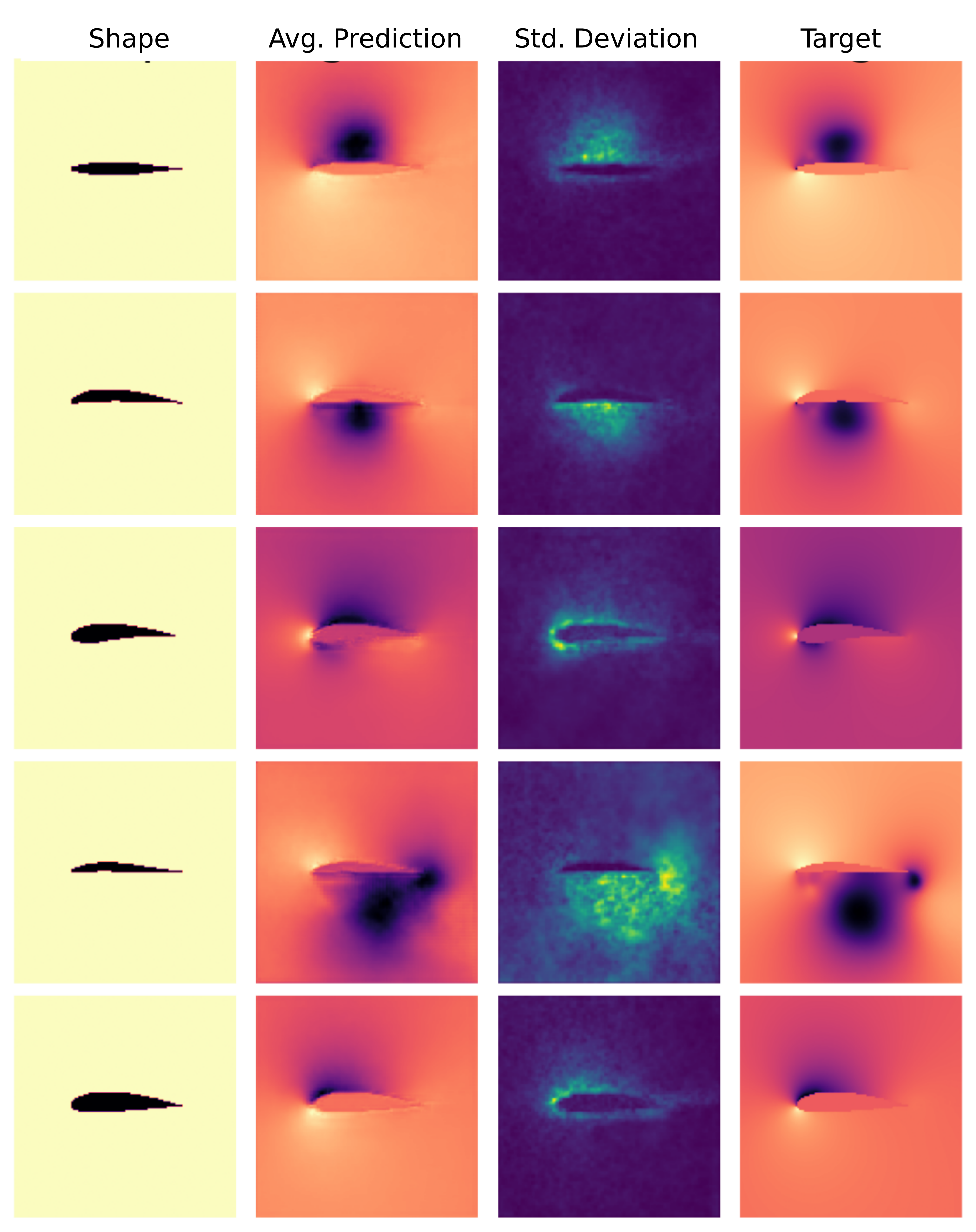}
    \caption{BNN with spatial dropout (left) and conventional dropout (right) acting on unseen shapes. \mm{The first column shows the input airfoil shape, the second column the average BNN prediction for the pressure field, the third column the corresponding uncertainty distribution, and the last column the true target pressure field.} The predictions and the uncertainty distribution are sensible: {closer to the airfoil and for low pressure pockets, the network is more uncertain.}}
    \label{fig:new_shapes}
\end{figure*}

\nt{The values for the BNN} in table \ref{tab:RansPerformance} are computed as follows: For every input sample, 20 forward passes are performed, and the average per-cell prediction is computed. This is illustrated in figure \ref{fig:spatial_vs_normal_repeated}, where the target is shown in the first column, followed by 5 of the 20 predictions. To the right, the averaged prediction is shown, and the last column corresponds to the uncertainty field, which is computed as the standard deviation per cell over the 20 predictions. The first row in figure \ref{fig:spatial_vs_normal_repeated} illustrates the forward pass for a BNN with spatial dropout, and the second row for a BNN with conventional dropout. Figure \ref{fig:spatial_vs_normal_repeated} also serves as an example of what has been observed generally when contrasting spatial dropout with conventional dropout: While the spatial dropout implementation leads to smooth large-scale variations across the repeated predictions, conventional dropout provides blurry, \nt{per-cell} noise. The resulting average predictions, however, are both smooth and similar to each other. The standard deviation, in contrast, again differs in scale and smoothness of the variation, even though the general location is similar and sensible for both implementations.
This is shown in figure \ref{fig:new_shapes}, where 5 unseen airfoil shapes (first column) are shown together with the corresponding average prediction (second column), uncertainty distribution (third column) and target (fourth column), both for spatial and conventional dropout. 

Across all shapes, the average prediction matches the target distribution closely, and the bulk of uncertainty is located at more turbulent regions close to the airfoil. The full test set with all unseen shapes is plotted in the appendix (figures \ref{fig:alltest_spatialdo_001} and \ref{fig:alltest_do_01})), together with the MAE-uncertainty relation per test sample (figure \ref{fig:mae_unc_persample}).

\subsection{Perturbed buoyancy-driven Navier-Stokes flow}
By perturbing Navier-Stokes simulations in a controlled manner, we create a supervised multimodal learning setup. We deploy a flipout-BNN and investigate if the multimodal nature \mm{inherent in the data-generating process can be captured by the stochastic predictions of the network. Our results show that more chaotic trajectories cause very significant difficulties for the trained BNNs. To investigate to what extent the BNN uncertainty allows us to distinguish unsuccessful cases from well performing ones, we relate the network's uncertainty to its predictive performance.}

\textbf{Setup.} 
We leverage the simulation framework \textit{PhiFlow} \citep{holl_learning_2020} to simulate time sequences of incompressible Navier-Stokes flows with a Boussinesq model for buoyancy forces. We simulate 2d trajectories on a $64\times64$ computational grid. 
For each trajectory, we add an inflow location with fixed $y-$ and random $x-$position and a buoyancy factor of $0.2$. 
We simulate 64 time steps and add noise to the velocity profile in every simulation step. Hence, the solver receives an already perturbed input and performs the solver step on noisy data. For large perturbations, this can lead to very chaotic trajectories, as shown via an advected marker field in figure \ref{fig:Perturbed-NS-example}. Formally, we can obtain a state $\mathbf{s}_{i}^{t}$ of trajectory $t$ at simulation time $i$ by repeatedly applying a Navier-Stokes simulation operator $\mathcal{P}$ and a perturbation operator $\mathcal{Q}$:
\begin{equation*}
    \mathbf{s}_{i}^{t}=\mathcal{P}\mathcal{Q}\mathbf{s}_{i-1}^{t}=\left(\mathcal{P}\mathcal{Q}\right)^{i}\mathbf{s}_{0}^{t}
\end{equation*}
where the state $\mathbf{s}_{i}^{t}=(\mathbf{d}_{i}^{t},\mathbf{v}_{i}^{t})$ with $\mathbf{d}_{i}^{t}$ denoting the marker field and $\mathbf{v}_{i}^{t}$ the velocity fields. \mm{The fluid model defining $\mathcal{P}$ is provided in the appendix \ref{app:perturbedNS} together with a formal description and a visualization (figure \ref{fig:perturbatonOperator}) of the perturbation operator $\mathcal{Q}$.}
The learning goal in this setup is to predict the velocity profiles advanced by $n$ simulation steps from the current marker profile, i.e. minimizing
\begin{equation}
    \E_{\vw\sim q_{\vtheta}} \left[ \sum_{i=1}^{N-n}\sum_{t=1}^{T}\norm{{f}(\mathbf{d}_{i}^{t}|\vw)-\mathbf{v}_{i+n}^{t}}_{1} \right] - \frac{1}{\lambda}\KL 
\end{equation}
with respect to $\mathbf{\theta}$.
\nt{In the following experiments, we use $T=100$ trajectories with $N=64$ frames each and an offset of $n=10$.}
We monitor uncertainty estimation and predictive performance over \mm{the range of noise loads $[0.,0.9]$.}
We deploy a U-Net with flipout layers in the decoder part and train it with the \textit{RMSprop} optimizer and a learning rate of 0.0014. 

\begin{figure}[t!]
    \centering
    \includegraphics[width=.5\textwidth]{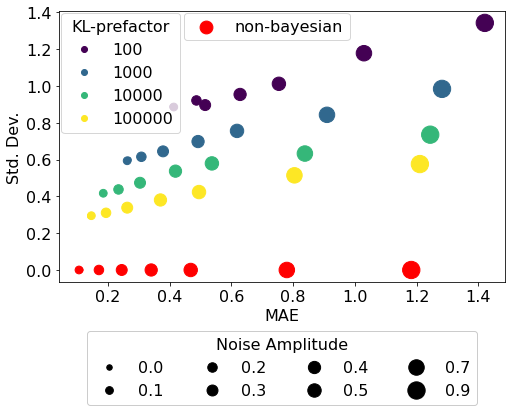}
    \caption{Uncertainty vs. mean absolute error for perturbed buoyancy-driven Navier-Stokes data. The performance of the conventional, non-Bayesian network is shown in red.}
    \label{fig:Uncert_vs_MAE_Phi_NPF}
\end{figure}

\textbf{Results.}
In figure \ref{fig:Uncert_vs_MAE_Phi_NPF}, a fine-grained analysis of the relation between mean absolute error and standard deviation is shown for the perturbed buoyancy driven Navier-Stokes data. Each dot represents the performance of an individual BNN, trained with a certain $\lambda$ value (indicated by the color of the dot) on data with a certain noise level (indicated by the size of the dot). The noise level ranges from unperturbed simulations (noise load 0.) to strongly perturbed simulations (noise load 0.9). The latter corresponds to very chaotic trajectories (an example is shown in figure \ref{fig:Perturbed-NS-example} in the appendix). In such cases, the network is typically not capable of reasonably approximating the target distribution. In figure \ref{fig:Uncert_vs_MAE_Phi_NPF} the transition from small $\lambda$ values (i.e. networks for which the KL term has larger weight) towards conventional, non-Bayesian networks with increasing $\lambda$ values is \mm{clearly visible}: Across all noise levels, the non-Bayesian network (red) is performing best in terms of mean absolute error. For a given noise level, larger values of $\lambda$ \mm{furthermore} always imply MAE values closer to the conventional network's performance. Like in the RANS flow case, BNNs are hence capable of obtaining predictive performance similar to their non-Bayesian counterparts (for large KL-prefactors), even though in this example they cannot outperform them. Also, the MAE-uncertainty relation is sensible: For each $\lambda$, the uncertainty is an increasing function of the MAE. The exact relation, however, depends on $\lambda$ itself. It is close-to-linear for small KL-prefactors (for e.g. $\lambda=100$, a linear regression yields a slope of 0.47) and becomes sub-linear for larger prefactors. 
A mapping from uncertainty to MAE is hence in principle possible. 

\begin{figure*} [tb!]
    \centering
    \includegraphics[width=1.\textwidth]{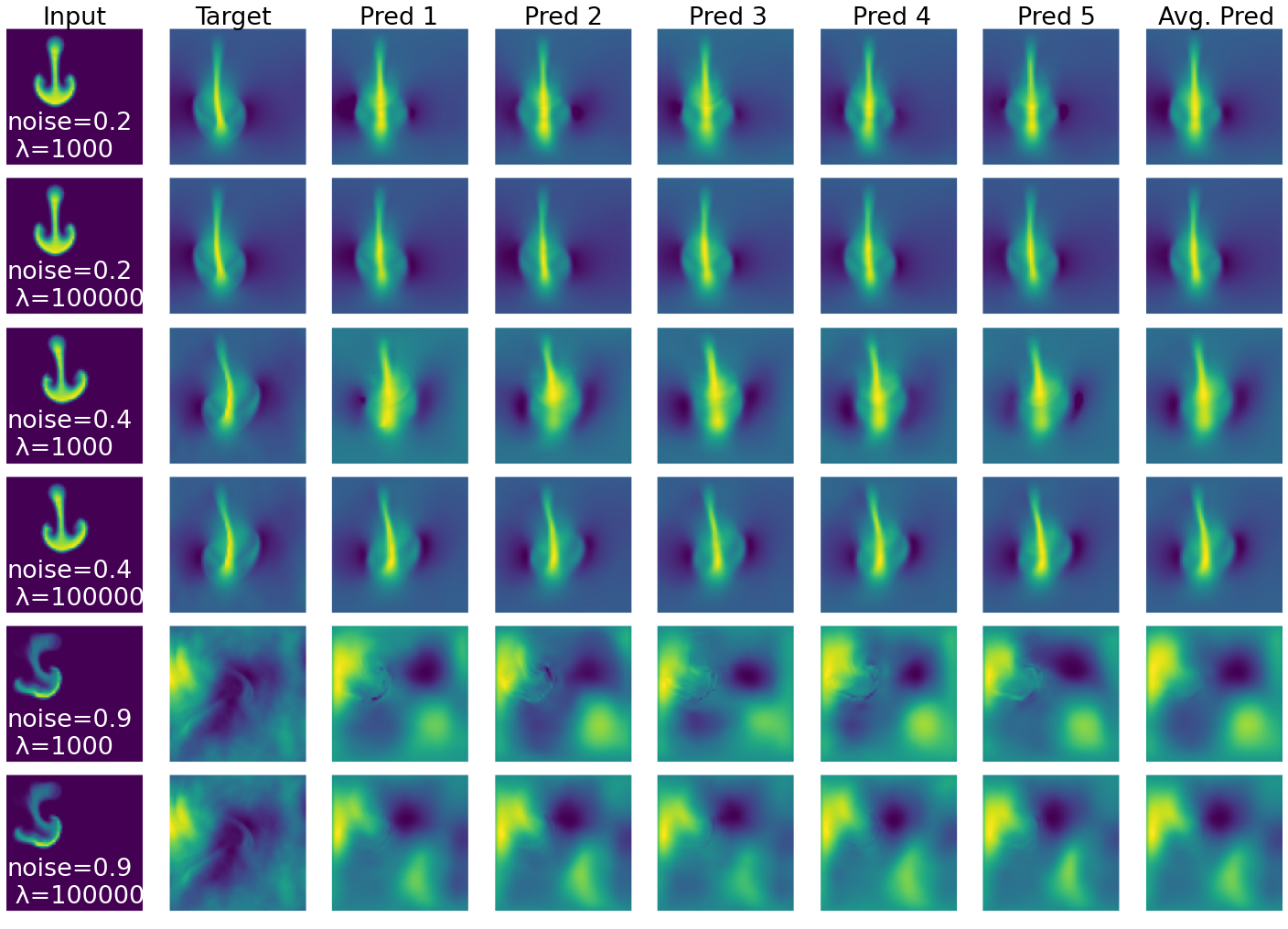}
    \caption{Repeated samples for different noise-per-frame training setups with $\lambda\in\{10^{3}, 10^{5}\}$ and noise amplitude $\in\{0.2, 0.4, 0.9\}$.}
    \label{fig:NPF_RepeatedSamples}
\end{figure*}

\nt{\textbf{Multimodality.}}
The extent to which the stochastic BNN predictions can capture the multimodal nature of the data can be seen qualitatively in figure \ref{fig:NPF_RepeatedSamples}. Each row shows repeated predictions of a network (trained with the noise level and $\lambda$ value indicated in the first column) for a certain input. For a given noise level, larger values of $\lambda$ lead to more precise predictions and fewer variations across repeated samples. In row 3 and 4 of figure \ref{fig:NPF_RepeatedSamples}, for instance, the model with larger $\lambda$ value approximates the target distribution better, but shows very little variation. The model with smaller $\lambda$ value shows more variation, but cannot approximate the detailed structures of the target distribution. Large noise levels, with examples shown in row 5 and 6, lead to very chaotic trajectories and cannot be approximated well at all. \mm{It is reassuring to see that the uncertainty for these very difficult learning tasks is significantly larger than for low-noise cases that were predicted reliably.}

\nt{Across all noise levels, smaller $\lambda$ values lead to more variations. However, they do} not capture large, physically sensible multimodal solutions (e.g. the plume being twisted to the left instead of the right). Instead, for a given noise load the target is just approximated worse, with mostly small-scale variations in the predictions. Hence, obtaining sensible multimodal solutions with the naive approach of a BNN as surrogate model to a fluid solver was not successful with the proposed setup. However, we will revisit this goal in the next experiment and show that multimodal solutions can be achieved when BNNs are deployed via a solver-in-the-loop training.

\subsection{Plasma Turbulence Simulations} 
In our third experiment, we consider a variation of the previous Navier-Stokes simulations: the simulation of \nt{drift wave turbulence} in plasma transport. 
We use the two-dimensional Hasegawa-Wakatani system, which is a simplified, yet powerful coupled set of equations relating the number density field ${n}$ with the electrostatic potential $\phi$ and its vorticity $\Omega$.
We leverage a differentiable implementation of the Hasegawa-Wakatani model \citep{greif_unpublished_2021}
and deploy a dropout BNN to work as a corrector function, as suggested for non-Bayesian networks and other fluid contexts by \citet{um_solver---loop_2021}. Our experiment shows that the BNN can not only significantly stabilize the simulations (like its non-Bayesian counterpart) but also successfully generate multimodal trajectories.

\textbf{Setup.} 
The \textit{solver-in-the-loop} setup from \citet{um_solver---loop_2021}, was demonstrated to have advantages in the setting of learned corrector functions for PDE solvers with regular, deterministic neural networks. 
\mm{Building upon this work, the} \nt{HW-solver in our experiment is likewise} realized such that it allows for gradient flow that supports backpropagation. During training, we simulate several predictor-corrector steps, compute losses with respect to a fine-grained pre-computed solution, and then backpropagate through the solver and corrector steps in order to update the corrector network parameters. Intuitively, this allows the network to explore the true, underlying physics and thus learn the highly non-linear nature of the errors. Furthermore, the corrector network ideally learns to correct its own behavior to reach a  steady state in the learning process. Formally, the learning problem can be written as minimizing 

\begin{equation*}
    \E_{\vw\sim q_{\vtheta}} \left[ 
    \sum_{i=0}^{n-1}\norm{\mathcal{C}(\mathcal{P}_{S}(\Tilde{\mathbf{s}}_{t+i})|\vw)-\mathcal{T}\mathbf{r}_{t+i+1}}^{2}_{2}
    \right] - \frac{1}{\lambda}\KL 
\end{equation*}
with respect to $\mathbf{\theta}$. $\mathcal{C}$ is the corrector network with weights $\vw$, $\mathcal{P}_{S}$ the solver, $\Tilde{\mathbf{s}}_{t+i}$ a simulation state at time $t+i$ (that has potentially already been corrected in previous steps), and $\mathcal{T}\mathbf{r}_{t+i+1}$ is the ground truth state the simulation is compared to.
Details for this setup are given in the appendix. Deploying a BNN as corrector network allows us to obtain a varied solution space for a given set of initial conditions, since a correction is then non-deterministic: Starting from the same initial frame, unrolling different trajectories is made possible by repeatedly applying the solver and the stochastic BNN corrector. 
In this setup, we deploy a ResNet with Bayesian dropout and enforce periodic boundary conditions by suitable padding. \nt{Details of this setup can likewise be found in the appendix.}

\begin{figure}[htb]
    \includegraphics[width=1.\columnwidth]{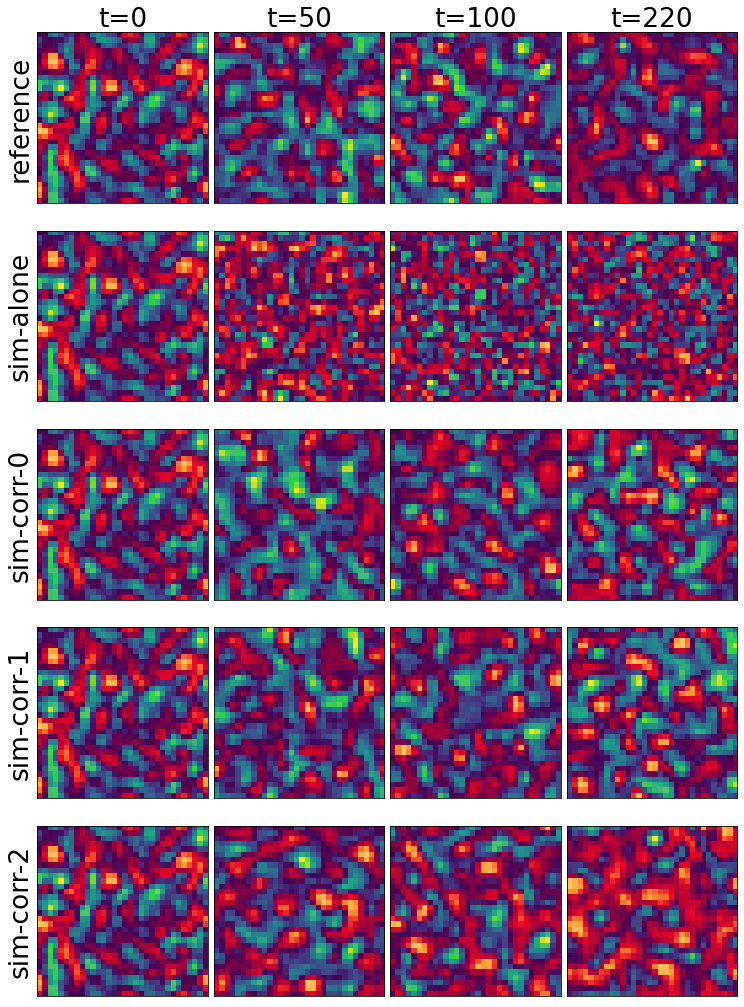}
    \caption{A BNN trained in the loop is able to stabilize turbulent Hasegawa-Wakatani simulations, as shown here with normalized $\phi$ fields. The stochastic nature of the corrections allows unrolling different trajectories from the same frame. The network was trained with spatial dropout and a rate of $0.01$.}
    \label{fig:phi_HW_SOL}
\end{figure}

\textbf{Results.} The simulations in this experiment are best inspected in the video available at \href{https://youtu.be/725ulH9JA-8}{https://youtu.be/725ulH9JA-8}. We furthermore illustrate them in figure \ref{fig:phi_HW_SOL}, where each row shows the frames of the temporal evolution of the $\phi$-field during a simulation. The reference in the first row was simulated in a grid of $128\times128$ cells, downsampled to a resolution of $32\times32$. This provides the ground truth solution. The second row (\textit{sim-alone}) shows the evolution of the trajectory of a simulation that was performed entirely in the source domain of size $32\times32$, without the BNN corrector. 
Hence, the first frame of all rows are identical, but the evolutions diverge over time. The \textit{sim-alone} simulation fails to maintain the turbulent state with large-scale structures that are clearly visible in row 1, producing mostly random noise towards the end of the simulation.
\mm{This is due to the low resolution discretization not resolving interactions with relevant, but unresolved, wavelengths which we aim to correct. When simulating with resolutions that are too coarse, energy accumulates at the grid scale, dominating the behavior and leading to the loss of all physical meaning.}

\begin{figure}[tb]
    \centering
    \includegraphics[width=1.\columnwidth]{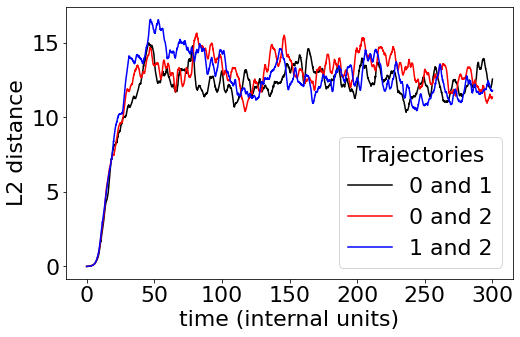}
    \caption{$L2$-distance between 3 BNN-corrected density trajectories: The initial correlation fades out after only about 30 internal time units (which is about 10 times the autocorrelation time).}
    \label{fig:L2_distance_HW_loop}
\end{figure}

Rows 3, 4 and 5 (\textit{sim-corr-0}, \textit{sim-corr-1} and \textit{sim-corr-2}) show 3 trajectories simulated in low resolution  that were unrolled with the BNN acting as corrector. In all three cases the \mm{scale of the} structures is preserved. Hence, the BNN was capable of learning suitable corrections such that
the turbulent state could be simulated in a stable manner. Importantly, the corrections in the last 3 rows were performed with the same trained network. Further, 
\mm{all three simulations start with the same initial conditions (see first column).} The different evolutions of the trajectories are caused by the stochastic nature of the corrections. It is interesting to see that the trajectories indeed differ from another significantly: From visible inspection, there is no clear correlation between the frames of the last 3 rows.
Figure \ref{fig:L2_distance_HW_loop} \mm{shows
the pairwise L2-distance between the 3 corrected trajectories as a function of simulation time and underlines this observation:} Since all trajectories start from the same frame, the distance is 0 initially, but increases over time. After about 30 internal time units, the distance stabilizes at the average L2 distance. This is about a factor of 10 larger than the autocorrelation time and hence shows, that after a short initial phase, the BNN indeed creates trajectories that are different to an extent where no correlation is measurable.
In the appendix, the same behavior as in figure \ref{fig:phi_HW_SOL} is illustrated for the $n$ and the $\Omega$ fields.

\section{Conclusions and Discussion}

We have presented a first study of BNNs deployed in a varied context of fluid simulations. We identified two use-cases of stochastic predictions, uncertainty assessment and multimodality, and investigated these in three complex setups with different variants of BNNs. 

We were able to show that BNNs can reach slightly superior performance when used as pure surrogate for the RANS-flow case. We argue that the increased performance is not caused by stronger regularization in BNNs. 
The BNN posterior can successfully capture many compelling but different solutions, which are combined in an ensemble-like manner during marginalization \citep{BayesianDLWilson}. We find qualitative differences in the uncertainty estimates when comparing spatial dropout to conventional dropout implementations, with spatial dropout providing smoother and larger variations. This is especially desirable, when realistic, individually sampled solutions are required, rather than the marginal prediction. We found the latter to be very similar for both spatial and conventional dropout. 

In the more challenging setup of perturbed Navier-Stokes flow, we could show that 
BNNs trained on more chaotic data 
consistently provided a larger uncertainty. This underlines the sensibility of the obtained uncertainty estimates. However, the BNNs were not able to capture large-scale multi-modalities of the flow data set for this case.

Finally, we employed a Bayesian \textit{solver-in-the-loop} framework to highly turbulent Hasegawa-Wakatani simulations. 
We could show that BNNs can successfully stabilize the simulations. 
Additionally, the stochasticity of the trained network can be leveraged to generate qualitatively different yet sensible trajectories from the same initial conditions. Especially for setups like the considered turbulent plasma physics simulations, this is very desirable. 
For such a case, the global statistical quantities, like average heat flux and particle transport,  are typically much more important than specific microscopic states. This poses a very interesting avenue for future work, namely whether the relevant, large-scale statistics can successfully be recovered by the BNNs.

\FloatBarrier

\bibliographystyle{icml2022}
\bibliography{references}


\onecolumn

\appendix

\section{Appendix}
\subsection{U-Net Architecture}
In the first and second experiment, we use a variant of the U-Net architecture as network model. Initially developed as a tool for image segmentation task, the U-Net architecture has shown to be a powerful tool across a wide variety of tasks and domains. It has a close-to-symmetric encoder-decoder-like architecture and uses skip-layer connections. In the contracting path (encoder), two $3\times3$ filters with strided convolutions followed by ReLU activation downsample each spatial dimension by $50\%$. At the same time, at every downsampling step, the number of feature channels is doubled. In the expansive part (decoder), an upsampling of the feature map increases the spatial resolution and is followed by a $2\times2$ convolution which halves the number of feature channels. Then, the resulting tensor is concatenated with the correspondingly cropped feature map from the encoder through a skip-layer connection. Finally, two $3\times3$ filters with ReLU activation are applied. In our implementation, we additionally apply batch normalization after the convolutional layers. In the first RANS-flow experiment, we use MC dropout, i.e. we apply (spatial) dropout after every layer. For the non-Bayesian network, we apply dropout only during training, whereas we extend it to the prediction phase for the Bayesian implementation. In the perturbed Navier-Stokes experiment, we deploy the U-Net as 'half-Bayesian' flipout network: We apply conventional layers in the encoder part, but TensorFlows Flipout layers in the decoder part of the network. 
\subsection{ResNet architecture}
In the final solver-in-the-loop experiment, we modify the ResNet that has been used in the original paper by \citet{um_solver---loop_2021}, consisting of 12 convolutional layers with kernel size 5 and 32 feature channels each. Again, we apply dropout after every layer and extend it to the prediction phase in order to obtain MC samples.
\FloatBarrier
\subsection{RANS-Flow}
\begin{figure*}[h]
    \centering
    \includegraphics[width=1.\textwidth]{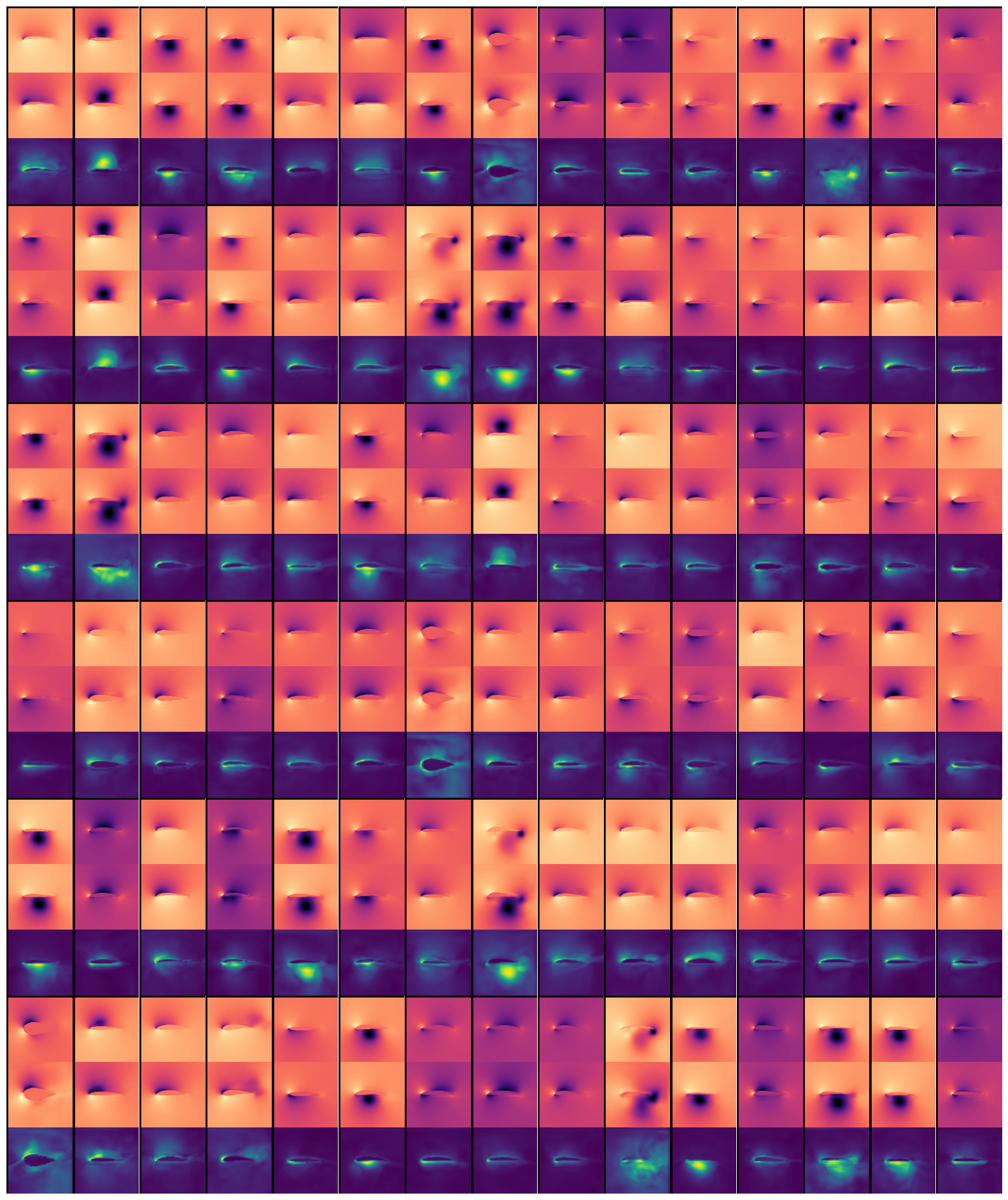}
    \caption{All test samples for a BNN with spatial dropout and dropout rate $0.01$. For each sample, the target is shown on top, the average prediction in the middle, and the corresponding uncertainty (in different color map) on the bottom.}
    \label{fig:alltest_spatialdo_001}
\end{figure*}

\begin{figure*}[h]
    \centering
    \includegraphics[width=1.\textwidth]{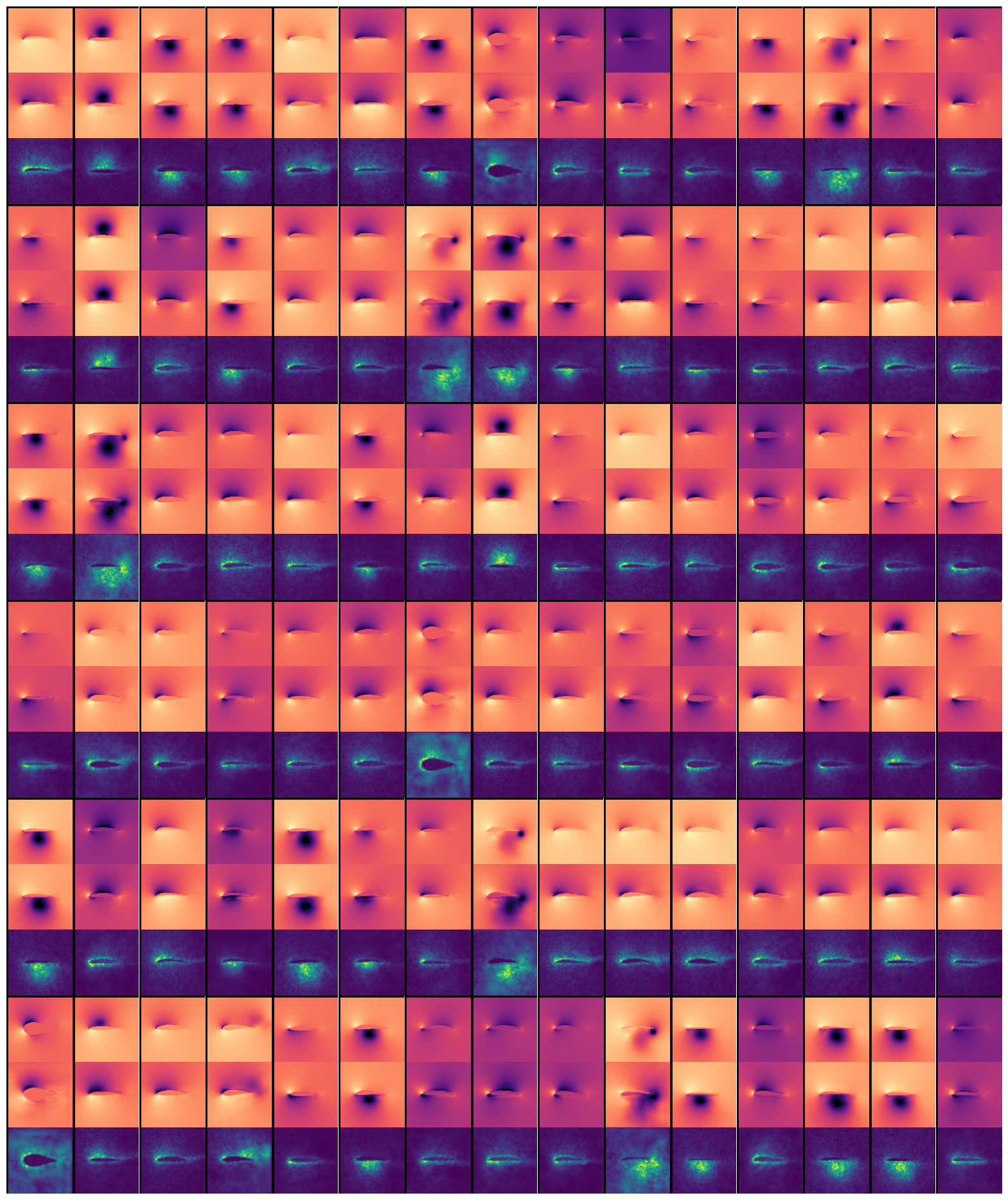}
    \caption{All test samples for a BNN with regular dropout and dropout rate $0.1$. For each sample, the target is shown on top, the average prediction in the middle, and the corresponding uncertainty (in different color map) on the bottom.}
    \label{fig:alltest_do_01}
\end{figure*}

\begin{figure*} [htb]
    \centering
    \includegraphics[width=1.\textwidth]{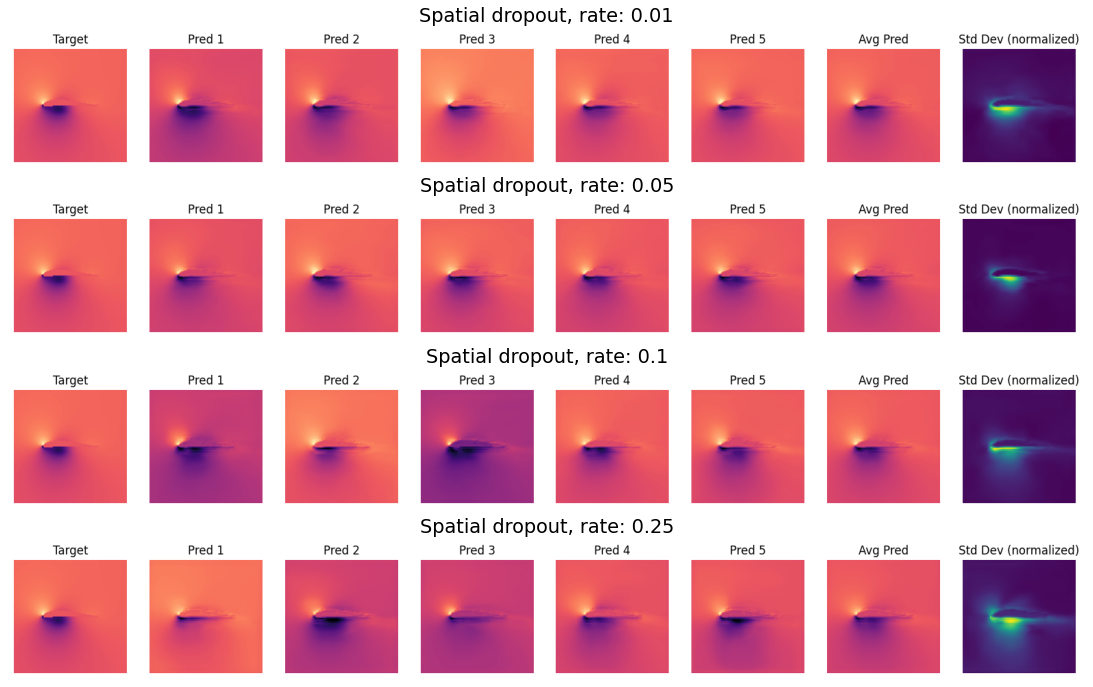}
    \caption{BNN samples (spatial dropout, different rates) for RANS flow.}
    \label{fig:repeated_all_spatialdo_rates}
\end{figure*}

\begin{figure*} [htb]
    \centering
    \includegraphics[width=1.\textwidth]{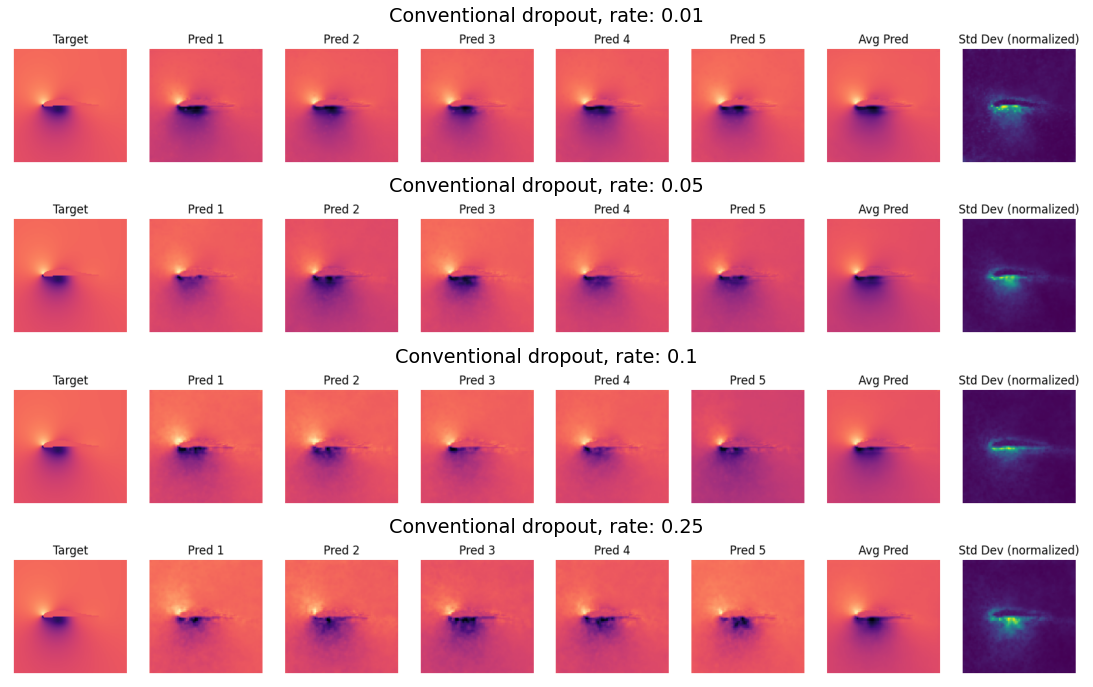}
    \caption{BNN samples (regular dropout, different rates) for RANS flow.}
    \label{fig:repeated_all_do_rates}
\end{figure*}

\begin{figure*}[!h]
        \centering
        \includegraphics[width=1.\textwidth]{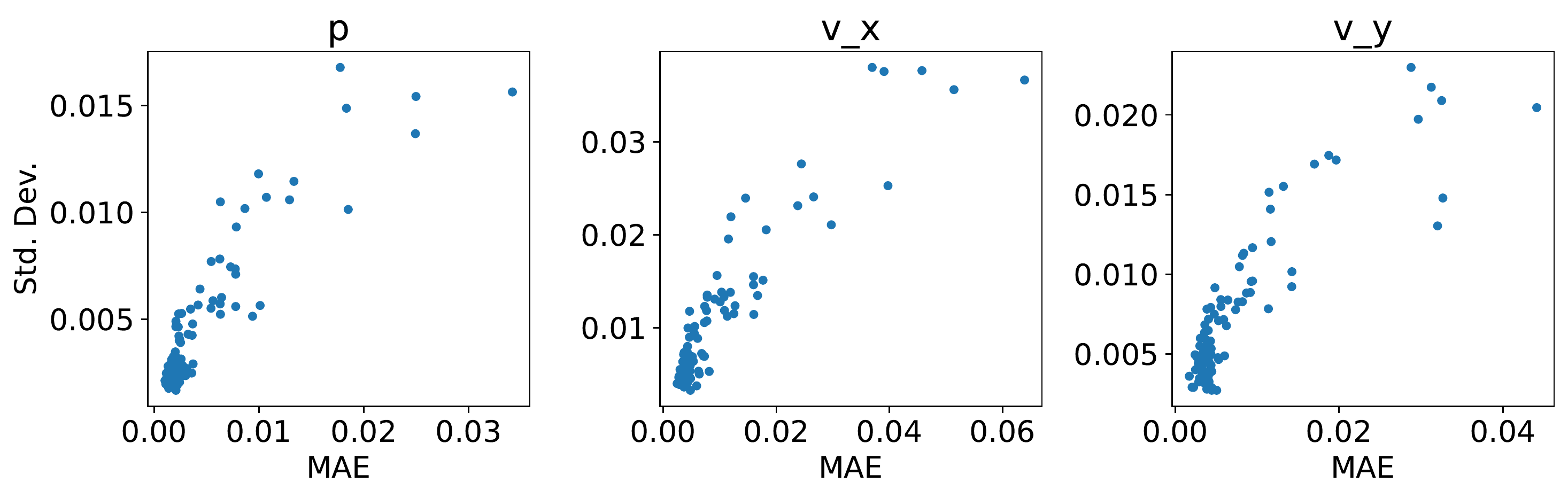}
        \caption{MAE vs. uncertainty per test sample for a BNN with spatial dropout rate 0.1: Larger uncertainty typically implies larger error.}        
        \label{fig:mae_unc_persample}
    \end{figure*}

For completeness, we provide all 90 test samples together with their marginal prediction and the corresponding uncertainty. Figure \ref{fig:alltest_spatialdo_001} illustrates the spatial dropout implementation with rate $0.01$, and figure \ref{fig:alltest_do_01} the conventional dropout implementation with rate $0.1$. Further, we show repeated samples for varying dropout rates, again for spatial dropout in figure \ref{fig:repeated_all_spatialdo_rates} and conventional dropout in figure \ref{fig:repeated_all_do_rates}. Figure \ref{fig:mae_unc_persample} shows the uncertainty-MAE relation of all test samples for a BNN with spatial dropout rate $0.1$.


\FloatBarrier
\subsection{Perturbed buoyancy-driven Navier-Stokes flow}\label{app:perturbedNS}
\begin{figure}[htb]
\centering
\includegraphics[width=.9\textwidth]{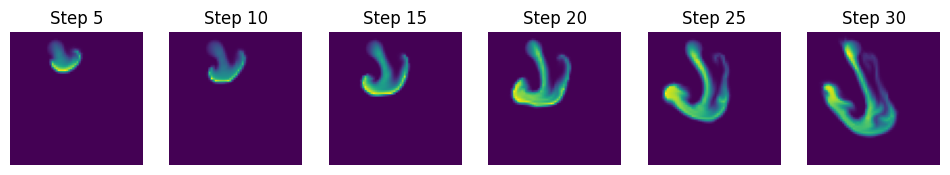}
\includegraphics[width=.9\textwidth]{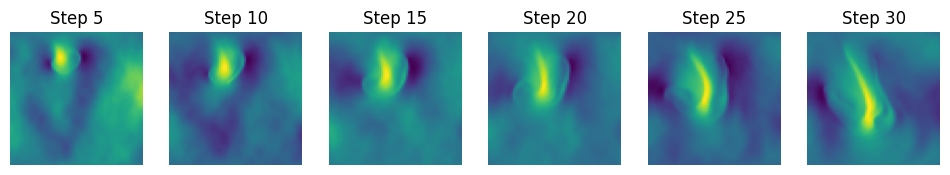}
\includegraphics[width=.9\textwidth]{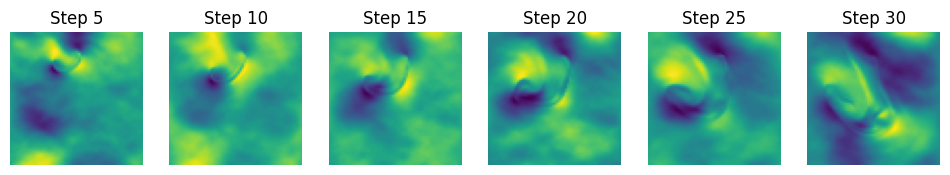}
\caption{Trajectories of marker field, $x-$component and $y-$component of velocity profiles for dynamically perturbed Navier-Stokes flow.}
\label{fig:Perturbed-NS-example}
\end{figure}

\begin{figure}
    \centering
    \includegraphics[width=.9\textwidth]{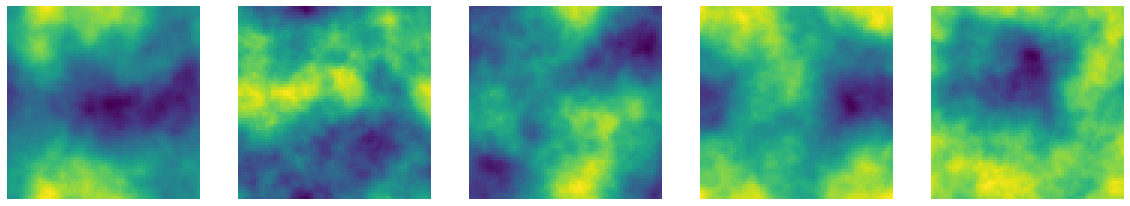}
    \caption{Samples of the noise fields which are added to the velocity fields by the perturbation operator $\mathcal{Q}$.}
    \label{fig:perturbatonOperator}
\end{figure}

\begin{figure}
    \centering
    \includegraphics[width=0.5\textwidth]{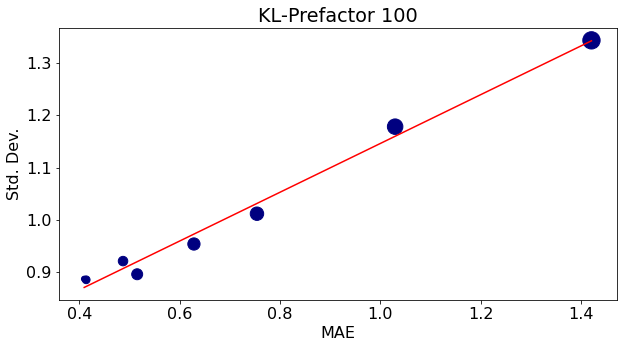}
    \caption{For $\lambda=100$, a linear regression between standard deviation and mean absolute error yields a slope of 0.47 and an intercept of 0.68.}
    \label{fig:linreg}
\end{figure}

The fluid model underlying the simulations are the Navier-Stokes equations in the boussinesq approximation:
$$\begin{aligned}
  \frac{\partial u_x}{\partial{t}} + \mathbf{u} \cdot \nabla u_x &= - \frac{1}{\rho} \nabla p 
  \\
  \frac{\partial u_y}{\partial{t}} + \mathbf{u} \cdot \nabla u_y &= - \frac{1}{\rho} \nabla p + \xi v
  \\
  \text{subject to} \quad \nabla \cdot \mathbf{u} &= 0,
  \\
  \frac{\partial v}{\partial{t}} + \mathbf{u} \cdot \nabla v &= 0 
\end{aligned}$$
with velocity field $\mathbf{u}$ and pressure $p$. $v$ is a scalar marker field indicating regions of higher temperature (or equivalently higher buoyancy force), and $\xi$ is a measure of the strength of the buoyancy force. Figure \ref{fig:perturbatonOperator} shows noise samples $\mathbf{n}$, like they are added to the velocity profiles by the perturbation operator $\mathcal{Q}$:
\begin{equation*}
    \mathcal{Q}\mathbf{s}_{i}^{t}=\mathcal{Q}(\mathbf{d}_{i}^{t},\mathbf{v}_{i}^{t})=(\mathbf{d}_{i}^{t},\mathbf{v}_{i}^{t}+\mathbf{n})
\end{equation*}
Figure \ref{fig:Perturbed-NS-example} shows samples of a perturbed trajectory, i.e. where $\mathcal{Q}$ was applied before every solver step. \mm{Figure \ref{fig:linreg} shows the result of a linear regression between standard deviation and mean absolute error for $\lambda=100$.}

\FloatBarrier
\subsection{Turbulent plasma simulations}

The Hasegawa-Wakatani system \cite{hasegawa_plasma_1983,wakatani_collisional_1984} is a simple fluid model relevant for the description of plasma turbulence. It assumes a constant magnetic field in $z$-direction, $\mathbf{B}=B\mathbf{\hat{z}}$ and can be written as 
\begin{equation}\label{eq:HaseWaka1}
       \frac{\partial}{\partial t}\nabla^{2}_{\bot}\Tilde{\phi}+(\mathbf{\hat{z}}\times\nabla_{\bot})\cdot\nabla_{\bot}\nabla_{\bot}^{2}\Tilde{\phi}=\alpha(\Tilde{\phi}-\Tilde{n})+D^{\phi}
\end{equation}
\begin{equation}\label{eq:HaseWaka2}
    \frac{\partial}{\partial t}\Tilde{n}+(\mathbf{\hat{z}}\times\nabla_{\bot})\cdot\nabla_{\bot}\Tilde{n}=\alpha(\Tilde{\phi}-\Tilde{n})+D^{n}
\end{equation}
where $x$, $y$ and $t$ are normalized to work as dimensionless spatial and temporal coordinates, $\Tilde{\phi}(x,y)$ is the normalized electrostatic potential, and $\Tilde{n}(x,y)$ the normalized density. \Eqref{eq:HaseWaka1} and \eqref{eq:HaseWaka2} are cross-coupled through the adiabaticity parameter $\alpha\propto\frac{TL_{n}}{n_{0}e^{2}c_{s}}$. The expressions $D^{\phi}$ and $D^{n}$ are non-physical terms added for numerical stability. Their role is to dissipate energy accumulating on grid scale through a so-called hyper-diffusivity. For details on the implementation, refer to \citet{greif_unpublished_2021}. $\nabla_{\perp}$ is to be understood as $(\partial x, \partial y, 0)$. Even though it might not be obvious at first glance, $\nabla_{\perp}^{2}{\phi}$ describes the fluid vorticity. This is because the fluid velocity can be approximated as the $\mathbf{E}\times\mathbf{B}$-drift and is hence $v_{D}\propto\mathbf{E}\times\mathbf{B}$. The vorticity can thus be written as
\begin{equation}
    \mathbf{\Omega}=\nabla\times \mathbf{v}_{D}\propto\nabla\times(-\nabla \phi\times\hat{\mathbf{z}})=\nabla^{2}\phi\hat{\mathbf{z}}.
\end{equation}
A more thorough introduction to the Hasegawa-Wakatani model, together with a detailed derivation of the equations, is available in chapter 2.5 of \cite{balescu_aspects_2005}, while \cite{camargo_resistive_1995} provides good insights into numerical experiments with the model.

\begin{figure*}[hbt]
    \centering
    \includegraphics[width=.9\textwidth]{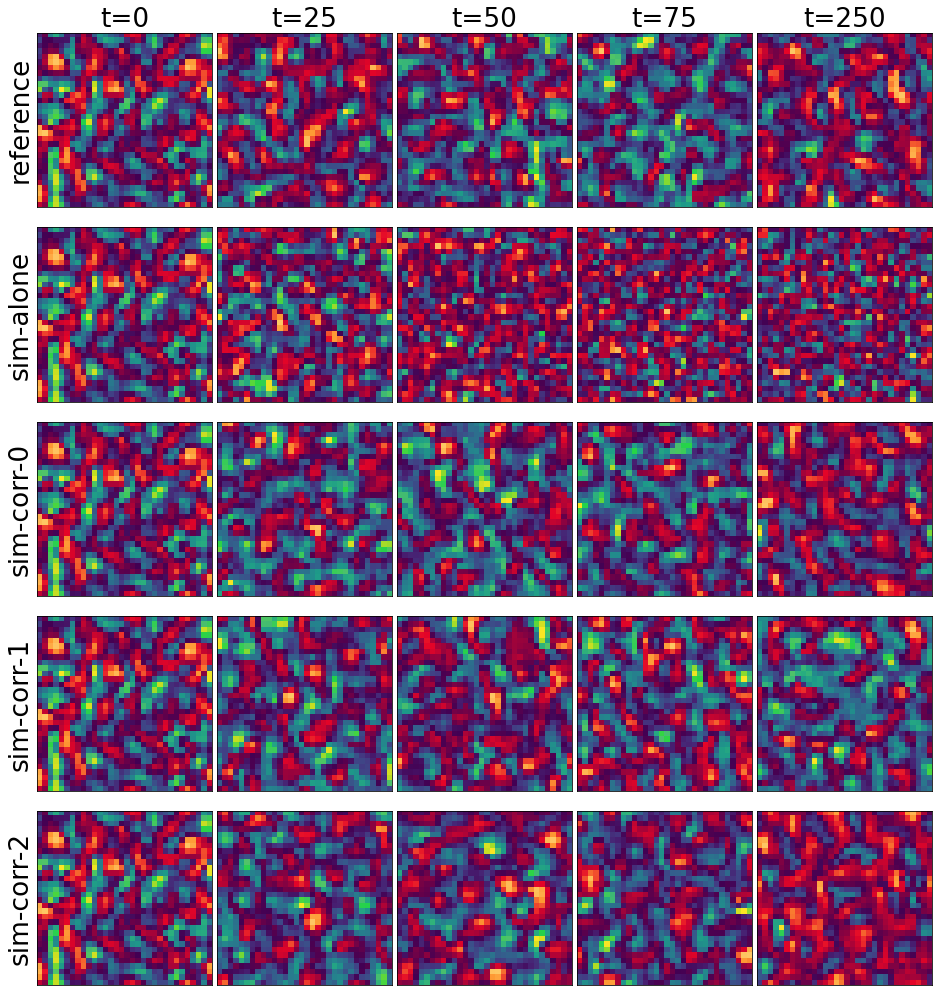}
    \caption{Density $n$}
    \label{fig:density_HW_SOL}
\end{figure*}

\begin{figure*}[hbt]
    \centering
    \includegraphics[width=.9\textwidth]{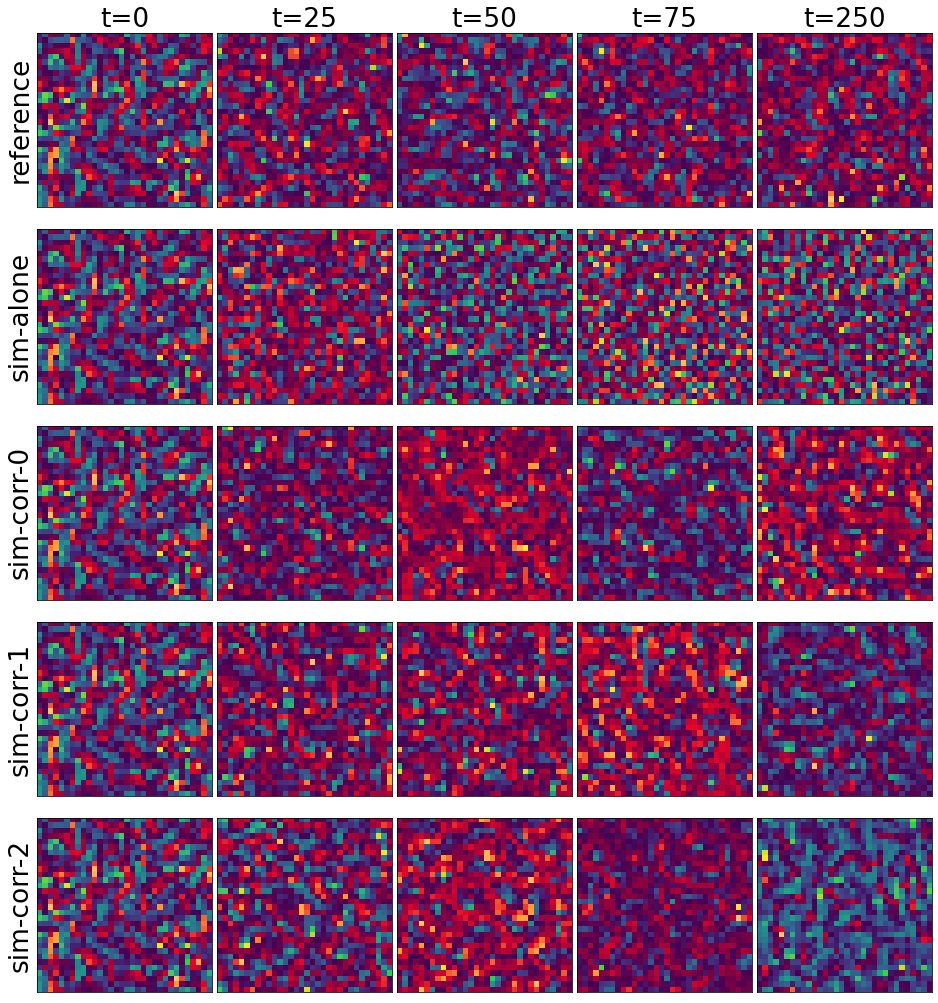}
    \caption{Vorticity $\Omega$}
    \label{fig:omega_HW_SOL}
\end{figure*}

\FloatBarrier

\subsection{Solver-in-the-loop Training Algorithm}\label{SOL-appendix}
The solver-in-the-loop setup can be formalized as follows. One considers two different discretization schemes of the same PDE $\mathcal{P}$: A fine reference discretization $\mathcal{P}_{R}$ with solutions $\mathbf{r}\in\mathcal{R}$ from the \textit{reference manifold} and a more coarse source discretization $\mathcal{P}_{S}$ with solutions $\mathbf{s}\in\mathcal{S}$. $\mathbf{r}$ and $\mathbf{s}$ are states at certain instances in time, but with different $\textit{spatial}$ resolution. Evolutions of the PDE consist then of a more exact reference sequence $\{\mathbf{r}_{t},\mathbf{r}_{t+\Delta t},...,\mathbf{r}_{t+k\Delta t},\}$ and a more coarse source sequence $\{\mathbf{s}_{t},\mathbf{s}_{t+\Delta t},...,\mathbf{s}_{t+k\Delta t},\}$, respectively. Furthermore, a mapping operator $\mathcal{T}$ is defined such that it transforms a phase space point from the reference manifold to the source manifold. 

In our case, where we consider the manifolds to differ only spatially, $\mathcal{T}$ can be seen as a downsampling operator, and we can write  $\mathbf{s}_{t_{0}}=\mathcal{T}\mathbf{r}_{t_{0}}$ for the initial state. A trajectory in the phase space, i.e. the evolution of a starting point $\mathbf{r}_{t}$, is obtained by evaluating $\mathcal{P}_{R}$ iteratively. A state after $k$ steps of equal step size $\Delta t$ is then $\mathbf{r}_{t+k}$. Importantly, $\mathcal{P}_{S}(\mathcal{T}\mathbf{r}_{t})\neq\mathcal{T}\mathbf{r}_{t+1}$. In other words, applying a solver step in the fine reference manifold and downsampling afterwards is not the same as downsampling first and applying the solver step in the coarse source manifold. The numerical error in the case where one applies the solver step in the coarse source manifold is typically larger, and this is precisely what we want to mitigate here. The learning goal is thus to train a correction operator $\mathcal{C}(\mathbf{s}|\mathcal{\theta})$, such that a corrected solution $\mathcal{C}(\mathbf{s})$ is closer to the downsampled solution than an unmodified state, e.g. in terms of the $L2$-norm:

\begin{equation*}
    \norm{\mathcal{C}(\mathcal{P}_{S}(\mathcal{T}\mathbf{r}_{   t_{0}}))-\mathcal{T}\mathbf{r}_{t_{1}}}<\norm{\mathcal{P}_{S}(\mathcal{T}\mathbf{r}_{t_{0}})-\mathcal{T}\mathbf{r}_{t_{1}}}
\end{equation*}

In our case, $\mathcal{C}(\mathbf{s}|\mathcal{\theta})$ is a neural network that receives a state $\mathbf{s}$ as input and corrects it to $\Tilde{\mathbf{s}}$. Further, repeated applications of the solver are denoted exponentially: 
\begin{equation*}
    \mathbf{s}_{t+n}=\mathcal{P}_{S}(\mathcal{P}_{S}(...\mathcal{P}_{S}(\mathcal{T}\mathbf{r}_{t})))=\mathcal{P}_{S}^{n}(\mathcal{T}\mathbf{r}_{t})
\end{equation*}

A fully corrected trajectory, where the correction is applied in every iteration, is then given by $\Tilde{\mathbf{s}}_{t+n}=(\mathcal{C}\mathcal{P}_{S})^{n}(\mathcal{T}\mathbf{r}_{t})$.
The solver-in-the-loop now leverages the differentiable physics pipeline that is available through \textit{PhiFlow} and integrates the solver into the training loop when training $\mathcal{C}$. Thus, the corrector $\mathcal{C}$ receives states $\Tilde{\mathbf{s}}$ that it has corrected previously and can backpropagate gradients through the solver steps. The objective function is thus

\begin{equation}\label{eq:SOL_objective}
    \argmin_{\theta}\sum_{i=0}^{n-1}\norm{\mathcal{C}(\mathcal{P}_{S}(\Tilde{\mathbf{s}}_{t+i})|\theta)-\mathcal{T}\mathbf{r}_{t+i+1}}^{2}_{2}.
\end{equation}

\begin{figure*}[!tb]
    \centering
    \includegraphics[width=.7\textwidth]{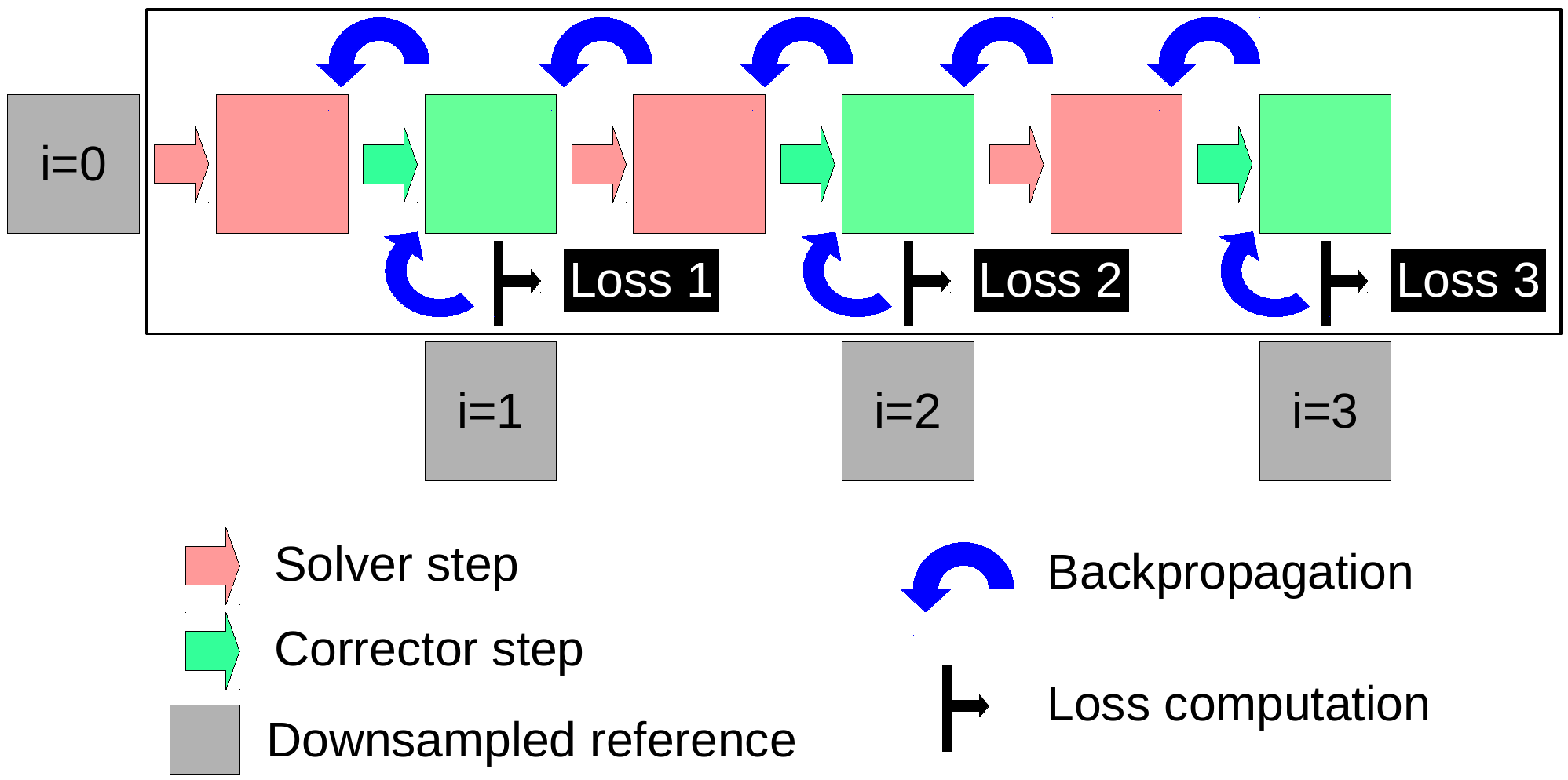}
    \caption{Sketch of the solver-in-the-loop with a look-ahead of $L=3$ and batch size $B=1$: Starting from a downsampled reference at $i=0$, solver and corrector are applied iteratively 3 times. Each corrected frame is compared to the corresponding downsampled reference frame and the respective losses are computed. Those losses are then backpropagated through the solver and the corrector in order to update the parameters of the corrector network.}
    \label{fig:SOL_flow}
\end{figure*}

The subtlety of $\eqref{eq:SOL_objective}$ is that the states $\Tilde{\mathbf{s}}_{t+k}$ themselves depend on the corrected function since they are computed via $\Tilde{\mathbf{s}}_{t+k}=(\mathcal{C}\mathcal{P}_{S})^{k}(\mathcal{T}\mathbf{r}_{t})$, leading to a recurrent optimization problem. In practice, one samples short simulation intervals of length $L\approx5$ instead of the full trajectory and uses batches of data. One starts with a collection of reference states, which are downsampled to the source domain. Then, each downsampled state evolves into a trajectory of size $L$ (the look-ahead) via recurrent application of $\mathcal{C}\mathcal{P}_{S}$. Each trajectory is compared to the corresponding reference trajectory, and the loss is computed. Backpropagation then allows to propagate gradients of the loss with respect to the network's parameters through all solver steps, like it is illustrated in figure \ref{fig:SOL_flow}. In pseudo-code, the optimization procedure can be written as 

\begin{algorithm}[bt]\label{algo:SOL}
\SetAlgoLined
\KwResult{Trains a corrector $\mathcal{C}(\theta)$ in the solver-in-the-loop framework.}
1. Inputs: A full, fine-grained reference trajectory. $\{\mathbf{r}_{0},\mathbf{r}_{1},...,\mathbf{r}_{N}\}$ \\
2. Set learning rate $\eta$, look-ahead $L$, batch size $B$ and initialize $\bm{\theta}$ randomly.\\
\While{$\bm{\theta}${ not converged}}
{Set $\Delta\mathbf{\theta}\longleftarrow 0$.\\
Sample a set $S$ of B random integers from the interval $[0,N-L]$.\\
\For{each $j$ in $S$}{
Obtain initial state in coarse-grained source domain by downsampling: $\mathbf{s}_{j}=\mathcal{T}\mathbf{r}_{j}$.\\
Compute L-look-ahead trajectory by applying solver and corrector iteratively and summarize to loss \\
\begin{equation}
    \mathcal{L}=\sum_{i=0}^{L-1}\norm{\mathcal{C}(\mathcal{P}_{S}(\Tilde{\mathbf{s}}_{j+i})|\theta)-\mathcal{T}\mathbf{r}_{j+i+1}}^{2}_{2}.
\end{equation}
Compute derivative of loss w.r.t. $\mathbf{\theta}$ by backpropagation through solver:\\
$\Delta \bm{\theta}\longleftarrow\Delta \bm{\theta}+\frac{\partial}{\partial \bm{\theta}}\mathcal{L}$.
}
Update $\bm{\theta}\longleftarrow\bm{\theta}+\eta\Delta\bm{\theta}$.
 }
 \caption{Solver-in-the-loop training}
\end{algorithm}

\end{document}